# THE *β* LYRAE-TYPE ECLIPSING BINARY EG Cep: NEW BVRI PHOTOMETRY AND MODELLING

CHLOI VAMVATIRA-NAKOU, ALEXIOS LIAKOS, VASSILIOS MANIMANIS, PANAGIOTIS NIARCHOS

*Department of Astrophysics, Astronomy and Mechanics*
*National and Kapodistrian University of Athens*
*GR 157 84 Zografos, Athens, Greece*

*Abstract*. New BVRI CCD observations of the semi-detached eclipsing binary EG Cep are presented. The observed light curves are analyzed with the Wilson-Devinney program and new geometrical and photometric elements are derived. These elements are used to compute the physical parameters of the system in order to study its evolutionary status.

## 1. INTRODUCTION

EG Cep ($\alpha_{2000} = 20^h\ 15^m\ 56.83^s$, $\delta_{2000} = 76°\ 48'\ 35.71''$) is a semi-detached eclipsing binary star with spectral type of the primary component A5. The system was discovered by Strohmeier (1958) as an eclipsing system with orbital period 0.5446 days. Its light curves are classified as EB (β Lyrae-type) in the General Catalogue of Variable Stars (Kholopov et al. 1985-1988). According to the study of Kaluzny and Semeniuk (1984) it is a 'semi-detached' system with mass ratio $q = 0.45$-$0.50$. Several investigators have studied the period changes of the system. Erdem, et al (2005), after analyzing their photoelectric data, suggested that the main U-shape (O-C) variation could be explained by a mass transfer from the less to the more massive star, a possible mass loss from the system and/or by the presence of a low mass unseen third component. There is only one spectroscopic study of the system by Etzel & Olson (1993), who calculated a projected rotational velocity of the primary component. Their other spectroscopic details were not published.

## 2. THE OBSERVATIONS

The observations of EG Cep were made on 9, 12, 13 June and 21 August 2007 in I filter, on 19 and 21 August 2007 in B, V, R filters and on 23 August 2007 in V, R filters. The telescope used was the 40-cm Cassegrain telescope of the Observatory of the University of Athens equipped with a ST-8 XMEI camera.

The data reduction (differential photometry) was made with the program AIP4WIN (Berry & Burnell 2000). The stars GSC 4585-0413 and GSC 4585-0165 where used as comparison and check star, respectively. Using the method of Kwee & van Wöerden (1956) new times of minima were calculated from our observations and they are given in Table 1. For the phase diagrams we used the ephemeris calculated by Kreiner (2000): $MinI(HJD) = 2454264.55370 + 0^d.5446251 \times E$

*Table 1*

The new times of minima from our observation

| HJD of minima | Error | Type | Filter |
|---|---|---|---|
| 2454334.53760 | 0.00025 | II | B, V, R |
| 2454264.55370 | 0.00006 | I | I |
| 2454265.37058 | 0.00015 | II | I |

## 3. LIGHT CURVE SOLUTION

We analyzed our light curves with PHOEBE 0.28 program (Prša and Zwitter, 2005) which uses the 2003 version the Wilson-Devinney code (Wilson and Devinney, 1971; Wilson, 1979), assuming that there are no spots on the two components, since there is no appreciable difference between the two maxima of our light curves. We applied the code in mode 2 (detached binary), mode 4 (semi-detached binary, primary star fills its Roche lobe) and mode 5 (semi-detached binary, secondary star fills its Roche lobe).

Since no spectroscopic mass ratio is available, initial values for the mass ratio (q = 0.477) and the inclination (i = 85.7 °) were adopted from the paper of Erdem et al. (2005). The fixed value of the primary star temperature ($T_1$ = 8500 K) was obtained from its spectral type. The gravity darkening coefficients $g_1$, $g_2$ and the albedos $A_1$, $A_2$ of the primary and secondary components, respectively, were set to the theoretical values. The limb darkening coefficients $x_1$, $x_2$ were supplied by the code. The subscripts 1 and 2 are referred to the star being eclipsed at primary and secondary minimum, respectively.

The solutions converged in the modes 2 and 5 of the program. However, the potential parameter $\Omega_2$ in the mode 2 was smaller than $\Omega_{in}$, indicating that EG Cep is a semi-detached system. The parameters derived from the solutions in mode 5 are shown in Table 2 and the theoretical light B and V curves of mode 5 solution, along with the observed ones, are shown in Figure 1.

*Table 2*

Parameters of light curve solution in mode 5

| Parameter | Mode 5 | Parameter | Mode 5 |
|---|---|---|---|
| i (degrees) | 86.47 (1) | $x_2$ (B, V) | 0.7685*, 0.6323* |
| $T_1$ (K) | 8500 | $x_2$ (R, I) | 0.5254*, 0.4302* |
| $T_2$ (K) | 5577 (3) | $g_1$, $g_2$ | 1.0*, 0.32* |
| q (=$m_2/m_1$) | 0.4683 (10) | $A_1$, $A_2$ | 1.0*, 0.5* |
| $\Omega_1$ | 2.8721 (17) | $r_1$ (pole) | 0.410 |
| $\Omega_2 = \Omega_{in}$ | 2.8177* | $r_1$ (point) | 0.508 |
| $L_1/(L_1+L_2)$ (B) | 0.9104 (5) | $r_1$ (side) | 0.434 |
| $L_1/(L_1+L_2)$ (V) | 0.8716 (5) | $r_1$ (back) | 0.458 |
| $L_1/(L_1+L_2)$ (R) | 0.8570 (5) | $r_2$ (pole) | 0.295 |
| $L_1/(L_1+L_2)$ (I) | 0.8278 (3) | $r_2$ (point) | 0.423 |
| $x_1$ (B, V) | 0.5491*, 0.4817* | $r_2$ (side) | 0.307 |
| $x_1$ (R, I) | 0.3893*, 0.2974* | $r_2$ (back) | 0.340 |

* assumed

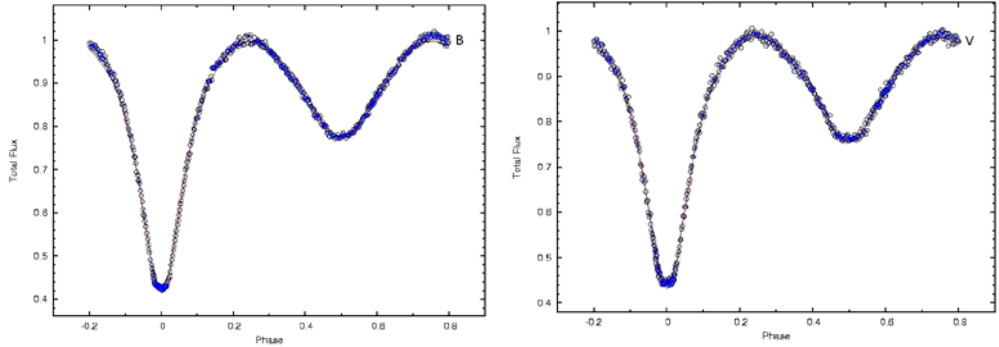

Fig. 1. Synthetic and observed B and V light curves of EG Cep.

## 4. ABSOLUTE ELEMENTS AND EVOLUTIONARY STATE

Since no double-line spectroscopy is available, we can estimate the absolute elements only by making assumptions about the mass of the primary star and using the mass ratio (q) obtained from the light curve solution. Assuming that the primary star has a mass of 1.86 solar masses, a typical value for a normal MS star of spectral type A5, we used the parameters derived from the light curve solution to compute the following absolute elements for EG Cep in solar units: $R_1 = 1.701 \pm 0.002$, $L_1 = 13.627 \pm 0.300$, $M_1 = 1.86$, $R_2 = 1.203 \pm 0.002$, $L_2 = 1.264 \pm 0.005$, $M_2 = 0.87$ and the bolometric absolute magnitudes: $M_{1,bol} = 1.912$ and $M_{2,bol} = 4.494$

The computed absolute elements are used in the Mass-Radius (M-R) diagram in order to speculate on the evolutionary status of the components of the system. The primary component is located between the TAMS and the ZAMS lines, so it is a slightly evolved star, whereas the secondary component seems to have evolved clearly beyong the TAMS line, which is a common fact among the near-contact systems.

## 5. SUMMARY AND CONCLUSIONS

Complete multi-colour light curves and CCD photometric analysis of EG Cep are presented for the first time. The results obtained from the light curve analysis show that the system of EG Cep is a β Lyrae-type semi-detached binary system whose secondary component fills its Roche lobe and so it is an evolved star beyong the TAMS line. The primary component is clearly detached from its Roche lobe.

*Acknowledgments*. The investigation was supported by the Special Account for Research Grants 70/4/5806 of the National & Kapodistrian University of Athens, Greece, and by UNESCO–BRESCE.